\documentclass[aps,prb,twocolumn,superscriptaddress]{revtex4}

\usepackage{amssymb}
\usepackage[dvips]{graphicx}
\usepackage{bm}
\usepackage{epsfig}
\usepackage[latin1]{inputenc}

\begin{document}


\title{Cd-doping effects in Ce$_2$MIn$_8$ (M = Rh and Ir) heavy fermion compounds}

\author{C. Adriano}
\email{cadriano@ifi.unicamp.br} \affiliation{Instituto de F\'isica
``Gleb Wataghin", Universidade Estadual de Campinas,
UNICAMP,13083-970, Campinas, São Paulo, Brazil.}

\author{C. Giles}
\affiliation{Instituto de F\'isica ``Gleb Wataghin", Universidade
Estadual de Campinas, UNICAMP,13083-970, Campinas, São Paulo,
Brazil.}

\author{E. M. Bittar}
\affiliation{Instituto de F\'isica ``Gleb Wataghin", Universidade
Estadual de Campinas, UNICAMP,13083-970, Campinas, São Paulo,
Brazil.}

\author{L. N. Coelho}
\affiliation{Instituto de F\'isica ``Gleb Wataghin", Universidade
Estadual de Campinas, UNICAMP,13083-970, Campinas, São Paulo,
Brazil.}

\author{F. de Bergevin}
\affiliation{European Synchrotron Radiation Facility, BP 220,
F-38043 Grenoble Cedex, France.}

\author{C. Mazzoli}
\affiliation{European Synchrotron Radiation Facility, BP 220,
F-38043 Grenoble Cedex, France.}

\author{L. Paolasini}
\affiliation{European Synchrotron Radiation Facility, BP 220,
F-38043 Grenoble Cedex, France.}

\author{W. Ratcliff}
\affiliation{NIST Center for Neutron Research, National Institute
of Standards and Technology, Gaithersburg, Maryland 20899, USA}

\author{R. Bindel}
\affiliation{NIST Center for Neutron Research, National Institute
of Standards and Technology, Gaithersburg, Maryland 20899, USA}

\author{J. W. Lynn}
\affiliation{NIST Center for Neutron Research, National Institute
of Standards and Technology, Gaithersburg, Maryland 20899, USA}
\author{Z. Fisk}
\affiliation{University of California, Irvine, CA 92697-4574, USA}

\author{P. G. Pagliuso}
\affiliation{Instituto de F\'isica ``Gleb Wataghin", Universidade
Estadual de Campinas, UNICAMP,13083-970, Campinas, São Paulo,
Brazil.}

\date{\today}

\begin{abstract}

Low temperature magnetic properties of Cd-doped Ce$_{2}$MIn$_{8}$
(M = Rh and Ir) single crystals are investigated. Experiments of
temperature dependent magnetic susceptibility, heat capacity and
electrical resistivity measurements revealed that Cd-doping
enhances the antiferromagnetic (AFM) ordering temperature from
\textit{T}$_{N}$ = 2.8 K (\textit{x} = 0) to \textit{T}$_{N}$ =
4.8 K (\textit{x} = 0.21) for Ce$_{2}$RhIn$_{8-x}$Cd$_x$ and
induces long range AFM ordering with \textit{T}$_{N}$ = 3.8 K
(\textit{x} = 0.21) for Ce$_{2}$IrIn$_{8-x}$Cd$_x$. Additionally,
X-ray and neutron magnetic scattering studies showed that Cd-doped
samples present below T$_{N}$ a commensurate antiferromagnetic
structure with a propagation vector $\vec{\varepsilon}$ =
$(\frac{1}{2},\frac{1}{2},0)$. The resolved magnetic structures
for both compounds indicate that the Cd-doping tends to rotate the
direction of the ordered magnetic moments toward the
\textit{ab}-plane. This result suggests that the Cd-doping affects
the Ce$^{3+}$ ground state single ion anisotropy modifying the
crystalline electrical field (CEF) parameters at the Ce$^{3+}$
site. Indications of CEF evolution induced by Cd-doping were also
found in the electrical resistivity measurements. Comparisons
between our results and the general effects of Cd-doping on the
related compounds CeMIn$_{5}$ (M = Co, Rh and Ir) confirms the
claims that the Cd-doping induced electronic tuning is the main
effect favoring AFM ordering in these compounds.
\end{abstract}



\maketitle


\section{INTRODUCTION}

Ce$_2$MIn$_8$ (M = Rh, Ir) are heavy-fermions compounds that
belong to the broader family Ce$_n$MIn$_{3n+2}$ (M = Co, Rh, Ir;
\textit{n} = 1, 2) where the occurrence of unconventional
superconductivity (USC) in various members has provided a great
opportunity to further study the relationship between magnetism,
USC and crystal structure. For the heavy-fermion superconductors
(HFS) in this family,  USC is believed to be mediated by magnetic
fluctuations\cite{Monthoux,Tuson} and their critical temperatures
seem to be related to the dimensionality of their crystal
structures.\cite{Hegger_PRL2000,Petrovic_JPCM2001,PG_PhysB_320_2002,PG_PRB2002,Pagliuso_PRB2001}
In fact, they are tetragonal variants of the cubic CeIn$_3$
structure with a layer of MIn$_2$ introduced between
\textit{n}-blocks of CeIn$_3$ along the \textit{c}-axis.
\cite{Hegger_PRL2000,Petrovic_JPCM2001}

Particularly interesting in these series is the tunability of
their ground state (GS) by pressure and/or doping which leads to
remarkable phase diagrams showing interplay between USC and
antiferromagnetism (AFM) and the presence of quantum critical
phenomena.~\cite{Tuson,PG_PhysB_320_2002,PG_PRB2002,Pagliuso_PRB2001,
Hegger_PRL2000,Petrovic_JPCM2001,pagliuso3,Ferreira_PRL2008,ericPRB,Zheeng,PG_PhysB_312_2002,Ramos_PhysB2005,Daniel,Pham_PRL2006,Nicklas,Urbano_PRL2007,Tuson2,Nicklas2}
For instance, applied pressure induces a high pressure
superconducting phase in the ambient pressure AFM members of this
family, \emph{n} = 1 (115) member CeRhIn$_5$ (\textit{T}$_N$ = 3.8
K) and \emph{n} = 2 (218) member Ce$_{2}$RhIn$_8$ (\textit{T}$_N$
= 2.8 K).~\cite{Tuson,Hegger_PRL2000,Tuson2,Nicklas2} The under
pressure properties of the single layer member CeRhIn$_5$ were
much more extensively studied. Not long ago, evidence for hidden
AFM order inside the SC state has been found in CeRhIn$_{5}$ under
pressure and magnetic field.\cite{Tuson} The importance of
simultaneous spin and charge fluctuations associated with a QCP
for the superconductivity in this material has been
shown.\cite{Tuson2}

Regarding the doping phase diagrams of the Ce$_n$MIn$_{3n+2}$ (M
=Co, Rh, Ir; \textit{n} = 1, 2) compounds, again the 115 members
have been the main focus of the reported studies in the
literature. For example, the reports on the CeRh$_{1-x}$(Co,
Ir)$_x$In$_5$ phase diagrams
~\cite{Pagliuso_PRB2001,Zheeng,PG_PhysB_320_2002} reveal
coexistence between AFM and SC for a large range of $x$.
Furthermore the critical temperatures of the superconducting
samples varies linearly with the \textit{c}/\textit{a} ratio of
the lattice parameters of the compounds \cite{PG_PhysB_312_2002}.
Many other chemical substitutions were made in the 115 materials
to contribute to the understanding of the microscopic tuning
parameter of their
GS.\cite{PG_PhysB_312_2002,pagliuso3,Ferreira_PRL2008,ericPRB,Ramos_PhysB2005,Daniel,Pham_PRL2006,Nicklas,Urbano_PRL2007}
La-doping on the Ce$^{3+}$ site for AFM members of the family
Ce$_n$MIn$_{3n+2}$ (M = Co, Rh, Ir; \textit{n} = 1, 2) were aimed
to describe the suppression the of the N\'{e}el temperature by
dilution and the effect of dimensionality on the percolation AFM
ordering.\cite{pagliuso3}

Indeed more revealing doping studies were made by the substitution
of In by Sn\cite{Ferreira_PRL2008,ericPRB,Ramos_PhysB2005,Daniel}
and Cd \cite{Pham_PRL2006,Nicklas,Urbano_PRL2007}. This is because
In, Cd and Sn are not isovalent. While Sn has one more
\textit{p}-electron than In, Cd has one \textit{p}-electron less,
and thus opposite electronic effects might be expected. However,
it was found that both elements suppress the SC state in CeMIn$_5$
for M = Co and Ir, while Cd-substitution tends to favor AFM
ordering but Sn-doping leads to its
suppression.\cite{ericPRB,Ramos_PhysB2005,Daniel,Pham_PRL2006,Nicklas,Urbano_PRL2007}
The microscopic mechanisms of this doping that tune the GS in the
CeMIn$_5$ compounds were attributed to the electronic tuning. It
is therefore important and elucidative to explore the behavior for
the other series in the Ce$_n$MIn$_{3n+2}$ (M = Co, Rh, Ir;
\textit{n} = 1, 2) family and determine to what extent the same
interpretation holds. In this regard, to study the \textit{n} = 2
member of the Ce$_n$MIn$_{3n+2}$ family, the Ce$_2$MIn$_8$ could
be particulary interesting since Cd and Sn doping are also
possible in the 218's and they present an intermediate step in
terms of dimensionality between the cubic CeIn3 and the 115
compounds.\cite{Hering_PhysB2008,Moreno1}

Recently, interest in the 218 family has increased due to the
discovery of the first HFS with a transition metal different from
Co, Rh or Ir, in the HF Ce$_{m}$MIn$_{3m+2}$ family.
Ce$_{2}$PdIn$_{8}$ was found to be an antiferromagnet
($\textit{T}_N$ = 10 K) and an ambient pressure SC
($T_{c}$~$\sim$~0.68~K). It has not yet been established whether
SC in this compound has an unconventional character or if it
coexists with the magnetism on a microscopic scale\cite{Ce2PdIn8}.

To further explore the microscopic effects for Cd-doping in the
Ce$_n$MIn$_{3n+2}$ (M = Co, Rh, Ir; \textit{n} = 1, 2) family we
present in this article the evolution of the low temperature
physical properties of the Ce$_2$MIn$_8$ (M = Rh, Ir) as a
function of the Cd-substitution. It is reported how the
\textit{T}$_{N}$ = 2.8 K and the magnetic structure (with
$\vec{\varepsilon}$ =
$(\frac{1}{2},\frac{1}{2},0)$)\cite{Bao_PRB2001} of Ce$_2$RhIn$_8$
evolves as a function of Cd-doping. Also the effects of Cd-doping
on the low temperature spin glass phase (\textit{T}$_g$ = 0.6
K)\cite{morris} of the HF Ce$_2$IrIn$_8$ are investigated. Our
results are compared to those found for the Ce115's compounds and
a discussion of the microscopic effect of the Cd in the GS of
these systems are also given.

\section{EXPERIMENTAL DETAILS}

Single crystalline samples of Ce$_2$MIn$_8$ (M = Rh, Ir) doped
with Cd were grown from Indium flux in which various amounts of Cd
were added to the flux,\cite{PG_PRB63_2001,Fisk_Handbook1989} so
that the nominal concentration of Cd can be defined by the Cd/In
ratio used in the growth. We have performed prompt gamma neutron
activation measurements on the instrument NG7 of the NIST Center
for Neutron Research of the National Institute of Standards and
Technology which reveal that the real Cd concentration in a given
crystal is about 13 \% of the nominal concentration for samples of
both series studied in this work. A detailed analysis of the
actual Cd concentration in Cd-doped CeMIn$_5$ made by Pham
\textit{et al}\cite{Pham_PRL2006} showed that the Cd concentration
in these crystals is approximately 10 \% of that in the flux from
which they were grown. In this work it will be used and labelled
in the figures the real Cd concentration measured by prompt gamma
neutron activation.

The tetragonal (P4/mmm) structure and unit cell parameters of all
samples were determined by X-ray powder diffraction. Within the
range of Cd-concentrations studied no appreciable changes in the
lattice parameters due the Cd-doping were observed.  Magnetization
measurements as a function of the temperature were performed using
a commercial superconducting quantum interference device (SQUID)
in the temperature range between 2.0 and 300 K. Specific heat and
electrical resistivity experiments were performed in a commercial
physical properties measurement system (PPMS) in the temperature
range between 0.3 and 10 K for the specific heat measurements and
between 1.9 and 300 K for the resistivity ones. The specific heat
set up uses a small mass calorimeter that employs a
quasi-adiabatic thermal relaxation technique, while the electrical
resistivity was measured using a low-frequency ac resistance
bridge and four-contact configuration. The samples were screened
and previously found to be free of surface contamination by
residual In-flux.

For the X-ray magnetic scattering (XMRS) experiments, a selected
crystal of Cd-doped Ce$_2$RhIn$_8$ with \textit{x} = 0.21 was
prepared with polished (00$l$) flat surface, and approximately 3
mm x 2 mm x 1 mm. The mosaic spread of the sample was found to be
$<0.05^o$ by a rocking curve ($\theta$ scan) on a four circle
diffractometer. XRMS studies were performed at the ID-20 beamline
at the European Synchrotron Radiation Facility (ESRF). The ID-20
X-ray source is a linear undulator with a 32 mm period. The main
optical components are a double Si(111) crystal monochromator with
sagittal focusing and two meridional focusing mirrors on either
side of the monochromator. The sample was mounted on a cryostat
(with a base temperature of 1.7 K using a Joule-Thomson stage),
installed on a four-circle diffractometer with the \textit{a}-axis
parallel to the beam direction. This configuration allowed
$\sigma$-polarized incident photons in the sample. In a second
run, the sample was mounted in a horizontal four circle
diffractometer with the (110) direction parallel to the beam
direction and this configuration allowed $\pi$-polarized incident
photons in the samples. During all measurements we performed a
polarization analysis, using LiF(220) crystal analyzer,
appropriate for the energy of Ce-L$_2$ absorption edge.

For the neutron magnetic scattering (NMS) experiments, selected
crystals of Cd doped Ce$_2$RhIn$_8$ and Ce$_2$IrIn$_8$ both with
\textit{x} = 0.21 of Cd were prepared with sizes of approximately
4 mm x 3 mm x 1 mm. NMS measurements were carried out on BT-9
triple axis neutron spectrometer of the NIST Center for Neutron
Research (NCNR). The samples were cooled in an ILL orange cryostat
(with a base temperature of 1.7 K), the horizontal collimators
used were 40-47-40-80. Neutrons with incident energy E = 35 meV
were selected using the (002) reflection of a pyrolytic graphite
monocromator, and filter to suppress the higher harmonics.
Uncertainties where indicated are statistical in nature and
represent one standard deviation.

\section{RESULTS AND DISCUSSIONS}

\subsection{Macroscopic properties of Ce$_2$(Rh,Ir)In$_{8-x}$Cd$_x$}

\begin{figure}
\begin{center}
\includegraphics[width=1.1\columnwidth,keepaspectratio]{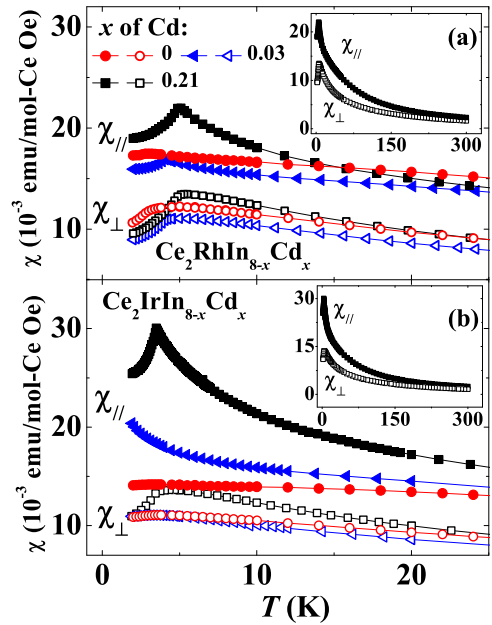}
\vspace{-2.0cm}
\end{center}
\caption{Temperature dependence of the magnetic susceptibility in
the low-\textit{T} region for Cd-doped (a) Ce$_2$RhIn$_8$ and (b)
Ce$_2$IrIn$_8$ single crystals (pure compounds (circles),
\textit{x} = 0.03 (triangles) and \textit{x} = 0.21 (squares)).
The magnetic field H = 0.1 T was applied parallel $\chi_{//}$
(closed symbols), and perpendicular $\chi_{\bot}$ (open symbols)
to the \textit{c}-axis. The insets in both panels show the
magnetic susceptibility as a function of the temperature from 2.0
K to 300 K for \textit{x} = 0.21 of Cd in both systems. 1 emu/(mol
Oe) = 4 $\pi$ 10$^-6$ m$^3$/mol.} \label{fig:Suscep_RhIr}
\end{figure}

Figure~\ref{fig:Suscep_RhIr} shows the temperature dependence in
the low-T region of the magnetic susceptibility measured for a
magnetic field H = 0.1 T applied parallel $\chi_{//}$ (closed
symbols), and perpendicular $\chi_{\bot}$ (open symbols) to the
\textit{c}-axis for pure and Cd-doped (a) Ce$_2$RhIn$_8$ and (b)
Ce$_2$IrIn$_8$ single crystals with Cd concentrations of 0.03
(triangles) and 0.21 (squares). The inset in each panel shows the
magnetic susceptibility as a function of the temperature from 2.0
K to 300 K for the samples with \textit{x} = 0.21. The results in
Fig.~\ref{fig:Suscep_RhIr}a for Cd-doped Ce$_2$RhIn$_8$ show AFM
order for \textit{T}$_N$ $\approx$ 5.0 K for \textit{x} = 0.21 of
Cd and \textit{T}$_N$ $\approx$ 4.0 K for \textit{x} = 0.03 of Cd.
The results in Fig.~\ref{fig:Suscep_RhIr}b for Cd-doped
Ce$_2$IrIn$_8$ show AFM order at \textit{T}$_N$ $\approx$ 3.8 K
for \textit{x} = 0.21 and no magnetic order to 2.0 K for
\textit{x} = 0.03. For both materials the magnetic susceptibility
is larger for a field applied along the $c$-axis, in agreement to
what was found for the compounds in the In-based
R$_{m}$M$_{n}$In$_{3m+2n}$ (M = Co, Rh or Ir, M = 1, 2) series
(except for R = Gd).
\cite{PG_PRB62_2000,Pagliuso_JAP2006,Serrano_PRB74_2006,Hieu_JMMM2007,Hudis_JMMM2006}
For the whole series the susceptibility is anisotropic and the
ratio $\chi_{//}$/$\chi_{\perp}$ taken at $\textit{T}_N$ is mainly
determined by the tetragonal crystalline electrical field (CEF)
and reflects to some extent the CEF anisotropy. The ratio remains
nearly constant as a function of Cd-doping for each family, for
example, for M = Rh the values of the ratio are roughly 1.6 and
1.5 for the samples with \textit{x} = 0.21 and 0.03, respectively.
For M = Ir the value of the ratio is nearly 2.2 and 2.1 for
samples with \textit{x} = 0.21 and 0.03, respectively. Fits from
the polycrystalline average of the magnetic susceptibility data
taken as $\chi_{poly}$=($\chi_{//}$+2$\chi_{\bot}$)/3 for
\textit{T} $>$ 150 K using a Curie-Weiss law yield an effective
magnetic moment $\mu_{eff}$ = 2.48(8) $\mu_{B}$ and a paramagnetic
Curie-Weiss temperature $\theta_{p}$ = - 44(1) K for M = Rh,
\textit{x} = 0.21, and $\mu_{eff}$ = 2.51(6) $\mu_{B}$ and
$\theta_{p}$ = - 42(1) K for M = Ir. The values of $\mu_{eff}$ and
$\theta_{p}$ do not change significantly as a function of the
Cd-concentration for both series in the range of concentration
studied.

\begin{figure}
\begin{center}
\vspace{-0.6cm}
\includegraphics[width=1.1\columnwidth,keepaspectratio]{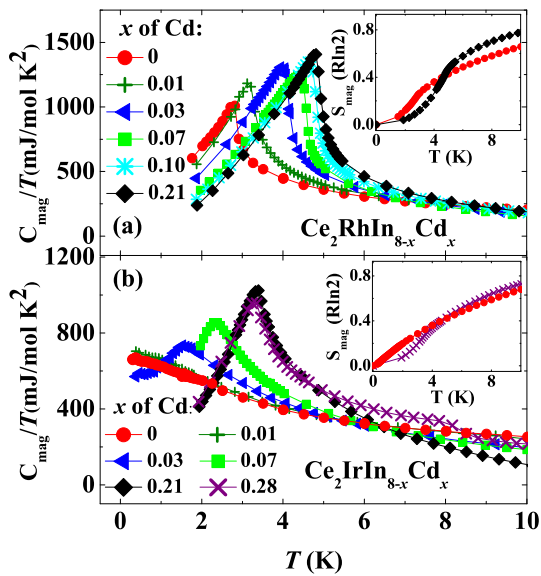}
\vspace{-2.5cm}
\end{center}
\caption{Magnetic specific-heat data (C$_{mag}$ = C$_T$ -
C$_{latt}$) divided by temperature as a function of temperature
for (a) Ce$_2$RhIn$_{8-x}$Cd$_x$ and (b) Ce$_2$RhIn$_{8-x}$Cd$_x$
single crystals for various Cd concentrations. The insets in each
panels show the magnetic entropy S$_{mag}$ for both
systems.}\label{fig:Cp_RhIr}
\end{figure}

Figure~\ref{fig:Cp_RhIr} shows the Cd-doping induced evolution of
the low temperature magnetic specific-heat C$_{mag}$ divided by
temperature for (a) Ce$_2$RhIn$_{8-x}$Cd$_x$ and (b)
Ce$_2$IrIn$_{8-x}$Cd$_x$ single crystals. To obtain the magnetic
specific-heat the lattice specific-heat contribution (C$_{latt}$)
was estimated from the specific-heat data of La$_2$RhIn$_8$ (not
shown) and subtracted from the total specific heat (C$_T$) of each
compound. The results of the specific-heat measurements for both
series clearly show an enhancement of the AFM order as a function
of Cd content in both systems. It is interesting to note that even
for the samples with larger quantity of Cd, the phase transition
anomaly is still a sharp and well defined peak, indicating a good
quality of the crystals and a reasonably homogenous distribution
of Cd concentration in the samples. More interesting is the case
of Ce$_2$IrIn$_{8-x}$Cd$_x$ in which the spin glass phase of the
pure compound seems to evolve to a long range AFM with increasing
doping even though doping typically tends to increase disorder.
This result suggests that the presence of a stronger Kondo effect
in Ce$_2$IrIn$_8$ is relevant to the formation of the frustrated
spin system in this compound.

The insets of both panels of Fig.~\ref{fig:Cp_RhIr} show the
corresponding magnetic entropy per mole Ce for the pure and
Cd-doped compounds in units of Rln2. For
Ce$_2$RhIn$_{8-x}$Cd$_{x}$ the recovered magnetic entropy for
\textit{x} = 0.21 at 10 K is about 0.75 Rln2 (closed diamonds)
that is a value larger than the recovered magnetic entropy seen
for the pure compound (0.65 Rln2, closed circles) at the same
temperature. In the case of Ce$_2$IrIn$_{8-x}$Cd$_{x}$ the entropy
of the Cd-doped sample is just slightly larger than the entropy of
the pure compound. These results suggest the magnetic entropy is
increasing as a function of the Cd concentration. On the other
hand, the recovered magnetic entropies for these compounds still
do not recover the Rln2 value expected for the entropy of the
ground state doublet, suggesting the presence of partly
compensated Kondo ordered moments.

\begin{figure}
\begin{center}
\vspace{-1.0cm}
\includegraphics[width=1.2\columnwidth,keepaspectratio]{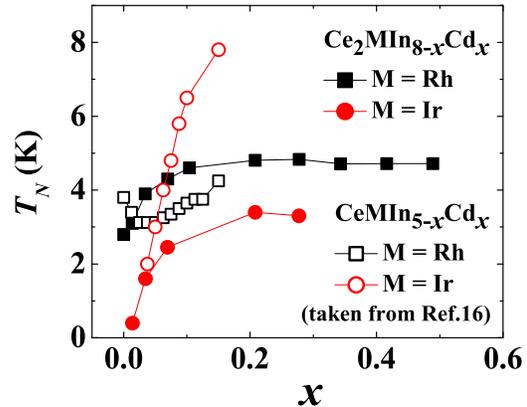}
\vspace{-2.5cm}
\end{center}
\caption{\textit{T}$_N$ as a function of Cd concentration for
Ce$_2$MIn$_{8-x}$Cd$_x$ (closed symbols) and CeMIn$_{5-x}$Cd$_x$
(open symbols) for M = Rh (squares) and Ir (circles). The 115's
results were taken from Ref.\onlinecite{Pham_PRL2006} for
comparison.}\label{fig:Tn}
\end{figure}

Figure~\ref{fig:Tn} summarizes the temperature-doping phase
diagrams resulting from heat capacity measurements for
Ce$_2$MIn$_{8-x}$Cd$_x$ (closed symbols), together with the
results found by Pham \textit{et al}. \cite{Pham_PRL2006} for
CeMIn$_5$ (open symbols) for M = Rh (squares) and Ir (circles).
For Ce$_2$RhIn$_{8-x}$Cd$_x$ one can observe the evolution of the
N\'{e}el temperature from 2.8 K for the pure compound to 4.8 K to
\textit{x} = 0.21 of Cd. For higher Cd-concentrations
\textit{T}$_N$ remains constant up \textit{x} = 0.35. In this
range of concentration there seems to be a saturation of the Cd
incorporation into the compound. For Ce$_2$IrIn$_{8-x}$Cd$_x$ one
can observe that a very small Cd concentrations (\textit{x} =
0.03) favors the establishment of long range AFM order when the
undoped \cite{morris} compound is a spin glass. The maximum value
of \textit{T}$_N$ for M = Ir is 3.8 K for \textit{x} = 0.21 of Cd.
For higher \textit{x} again a saturation of the Cd incorporation
in samples is observed.

Therefore the results of figure~\ref{fig:Tn} reveal that Cd-doping
favors AFM ordering in both CeMIn$_5$, (M = Co, Rh,
Ir)\cite{Pham_PRL2006} and Ce$_2$MIn$_8$ (M = Rh, Ir) HF families.
To further compare our results with those found by Pham \textit{et
al}. \cite{Pham_PRL2006} for Cd-doped Ce115, we notice that, for
CeRhIn$_5$, the Cd-doping first reduces the T$_N$ from 3.8 K for
undoped material to 3.0 K, with a flat minimum for Cd
concentration between 0.05 and 0.10. Higher Cd concentrations lead
to a N\'{e}el temperature increase up to 4.2 K for \textit{x}
$\sim$ 0.15. This behavior in Cd-doped CeRhIn$_5$ was suggested to
be connected with an incommensurate to commensurate magnetic
ordering evolution that can be induced by Cd-doping in CeRhIn$_5$
\cite{Pham_PRL2006}. In Cd-doped Ce$_2$RhIn$_8$ we found that
\textit{T}$_N$ is always increasing with the Cd concentration. The
fact that pure Ce$_2$RhIn$_8$ already shows just below
\textit{T}$_{N}$ = 2.8 K a commensurate magnetic structure (with
$\vec{\varepsilon}$ =
$(\frac{1}{2},\frac{1}{2},0)$)\cite{Bao_PRB2001} may explain the
difference between the two structures.

For CeIrIn$_5$ the Cd-doping initially  suppress the
superconductor state and for concentrations larger than 0.08 the
AFM state appear and the increase of Cd content results in an
increase of \textit{T}$_N$ from 2.0 K to around 8.0 K for
\textit{x} = 0.15. In the case of Cd-doped Ce$_2$IrIn$_8$ we found
that the spin glass state of the pure compound evolves to a long
range AFM state and \textit{T}$_N$ also increases as a function of
$x$, but the maximum \textit{T}$_N$ $\sim$ 3.8 K achieved is
smaller than the one found for Cd-doped
CeIrIn$_5$.\cite{Pham_PRL2006}

The favoring of AFM in the Cd-doped 115's was interpreted using a
Doniach-like scenario\cite{Doniach} where the Cd is inducing a
decrease of the local density of states at the Ce$^{3+}$ site
which tends to reduce the Kondo effect favoring the AFM
ordering\cite{Pham_PRL2006}. The behavior of T$_N$ found here for
the Cd-doped 218 compounds corroborates this interpretation since
the same Cd-doping effect would be expected to happen.

\begin{figure}
\begin{center}
\vspace{-2.0cm}
\includegraphics[width=1.3\columnwidth,keepaspectratio]{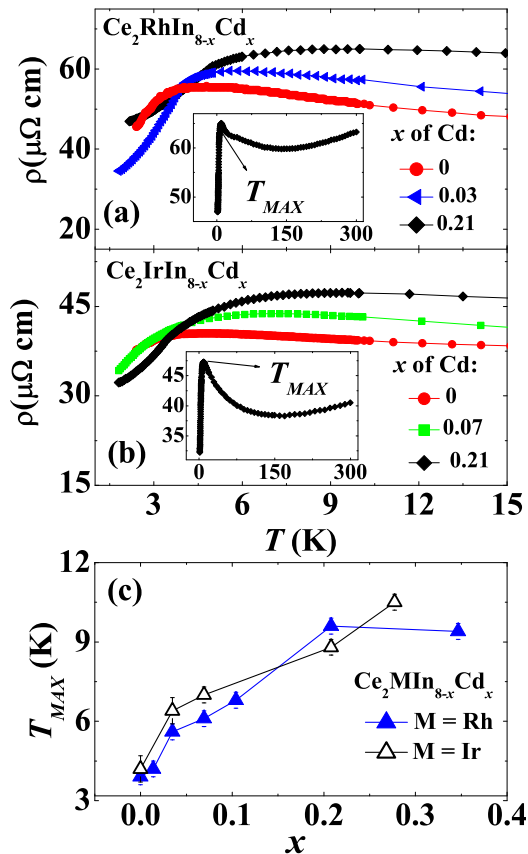}
\vspace{-1.0cm}
\end{center}
\caption{Temperature dependence of the electrical resistivity
$\rho$(\textit{T}) in the low-\textit{T} region for: (a)
Ce$_2$RhIn$_{8-x}$Cd$_x$ and (b) Ce$_2$IrIn$_{8-x}$Cd$_x$
compounds. The insets in each panels show $\rho$(\textit{T}) from
2.0 K to 300 K for \textit{x} = 0.21 for both series. The arrows
indicate the temperature \textit{T}$_{MAX}$ where the electrical
resistivity has a maximum; (c) \textit{T}$_{MAX}$ as a function of
the Cd concentration}\label{fig:Resist_RhIr}
\end{figure}

The temperature dependence in the low-\textit{T} region of the
electrical resistivity $\rho$(\textit{T}) is plotted in
Figure~\ref{fig:Resist_RhIr} panel (a) for M = Rh and Cd
concentrations of \textit{x} = 0.03 (triangles) and 0.21
(diamonds), and panel (b) for M = Ir and Cd concentrations of 0.07
(squares) and 0.21 (diamonds). The insets in
Fig.~\ref{fig:Resist_RhIr} (a) and (b) show the behavior of the
electrical resistivity for the entire range of temperatures for
the samples with \textit{x} = 0.21 of Cd for both systems. For all
crystals of the two series, the room-temperature value of the
electrical resistivity ranges between 40-80 $\mu\Omega$cm. Their
high-temperature data show a weak metallic behavior down to 150 K,
then the resistivity increases reaching a well-defined maximum at
a temperature \textit{T}$_{MAX}$, around 5.0-10 K (insets of the
panels (a) and (b)). The value of \textit{T}$_{MAX}$ from the
resistivity measurements for all compounds as a function of the Cd
content is plotted in Fig.~\ref{fig:Resist_RhIr}c for
Ce$_2$RhIn$_{8-x}$Cd$_x$ (open triangles) and
Ce$_2$IrIn$_{8-x}$Cd$_x$ (closed triangles).

The results of the resistivity at low-\textit{T} in
Fig.~\ref{fig:Resist_RhIr}a for Ce$_2$RhIn$_{8-x}$Cd$_x$ are
showing clear kinks around 2.8 K, 3.8 K and 4.8 K for \textit{x} =
0, 0.03 and 0.21 respectively, in good agreement with the values
obtained for specific-heat measurements for the same values of
\textit{x}. Also for Cd-doped Ce$_2$IrIn$_{8}$ in
Fig.~\ref{fig:Resist_RhIr}b one can observe the kink in the
resistivity measurements for \textit{x} = 0.21. The other curves
for pure and Cd-doped Ce$_2$IrIn$_{8}$ with \textit{x} = 0.07 do
not show obvious kinks at \textit{T}$_{N}$, perhaps because the
AFM ordering has not yet become long range for these
Cd-concentrations. The residual resistivity of the single crystals
of Cd-doped Ce$_2$MIn$_8$ (M = Rh, Ir) is roughly two orders of
magnitude larger than the residual resistivity of the CeMIn$_5$ (M
= Co, Rh, Ir) compounds. This behavior is consistent with the
typically larger chemical disorder of the 218 materials when
compared to that of the 115 materials.\cite{Eva1,Eva2}

The most prominent feature of the resistivity data is the
evolution of \textit{T}$_{MAX}$, which, as shown in
Fig.~\ref{fig:Resist_RhIr}c, is increasing as a function of the Cd
doping for both series. This behavior of \textit{T}$_{MAX}$ is not
expected within the Doniach-like scenario used to interpret the
behavior of \textit{T}$_{N}$ in these series.

From the point of view of the Doniach model, the decrease of a
local density of states induced by Cd-doping drives the system to
be more magnetic, i.e. the RKKY interactions are been favored
while the Kondo effect is decreased. Based on this model, the
\textit{T}$_{MAX}$ that is related to the Kondo energy scale
should shift to lower temperatures as a function of the Cd
concentration, in contrast to the observed results. This suggests
that the Cd-doping may have a second effect that produces the
increase of the maximum in the resistivity. To further understand
the role of Cd substitution in Ce$_2$MIn$_8$ we have performed
X-ray and neutron magnetic scattering experiments to determine the
low-\textit{T} magnetic structure with Cd-doping.

\subsection{Magnetic structure of Ce$_2$(Rh,Ir)In$_{7.79}$Cd$_{0.21}$}

Cd-doping perturbations in the AFM state were further explored
through X-ray and neutron  magnetic scattering experiments. For
these experiments we chose samples with the larger N\'{e}el
temperature for both series with Cd concentration of \textit{x} =
0.21 and  \textit{T}$_N$ = 4.8 K and 3.8 K respectively to M = Rh
and Ir.

The XRMS experiments were performed with the incident photon
energy at the Ce-L$_2$ absorption edge in resonant condition in
order to enhance the small signal from the AFM order of Ce$^{3+}$
ions below \textit{T}$_N$, and polarization analysis was performed
to more properly determine the moment direction. Magnetic peaks of
the Cd-doped Ce$_2$RhIn$_8$ with \textit{x} = 0.21 were observed
in dipolar resonant condition at temperatures below $\textit{T}_N$
= 4.8 K at reciprocal lattice points forbidden for charge
scattering and consistent with an antiferromagnetic structure with
propagation vector
$(\frac{1}{2},\frac{1}{2},0)$\cite{Cris_PhysB2009}. This is the
same propagation vector found by W. Bao et al. \cite{Bao_PRB2001}
for the pure compound, showing that Cd-doping is not changing the
propagation vector. This indicates that although the Cd-doping
enhances the magnitude of the average magnetic exchange between
neighboring Ce$^{3+}$ ions it does not affect the sign of the
relative magnetic interaction (and spin orientation) between them.

\begin{figure}
\begin{center}
\vspace{-1.0cm}
\includegraphics[width=1.5\columnwidth,keepaspectratio]{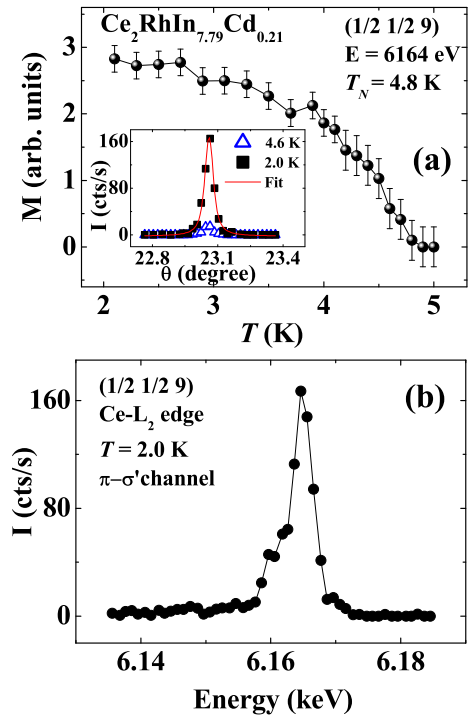}
\vspace{-1.0cm}
\end{center}
\caption{(a) Temperature dependence of the ordered magnetic
moment, proportional to square root of the integrated intensity,
of the $(\frac{1}{2},\frac{1}{2},9)$ magnetic reflection measured
with transverse $\theta$ scans in the temperature range between
\textit{T} = 2.0 and 5.2 K for Ce$_2$RhIn$_{7.79}$Cd$_{0.21}$. The
inset shows two $\theta$-scans of the same reflection for
different temperatures: 4.6 K (open triangles) and 2.0 K (closed
squares). The continuous line is a Voigt fit to the observed data
at T = 2.0 K. (b) Energy line shape of the
$(\frac{1}{2},\frac{1}{2},9)$ magnetic peak at T = 2.0 K for
 $\pi$-$\sigma$' polarization channel at the Ce-L$_2$ absorption
edge. The line is just a guide to the eye.}\label{fig:Temp_Rh}
\end{figure}

Figure~\ref{fig:Temp_Rh}a displays the temperature dependence of
the magnetic moment of the Ce$^{3+}$ ion, which is proportional to
the square root of the integrated intensities, of the
$(\frac{1}{2},\frac{1}{2},9)$ magnetic reflection at an incident
photon energy of 6164 eV (Ce-L$_2$ edge) and measured in
$\sigma$-$\pi$' polarization channel. A Voigt peak shape was used
to fit transverse $\theta$ scans (sample rotation) through the
reciprocal lattice points in order to obtain the integrated
intensities. The data were taken between \textit{T} = 2.0 K and
\textit{T} = 5.2 K while warming the sample. The inset shows the
experimental $\theta$-scans for two different temperatures,
\textit{T} = 4.6 K (open triangles) and \textit{T} = 2.0 K (closed
squares), where one can observe the decrease of the intensity as
\textit{T} approaches \textit{T}$_N$. The continuous line is a
Voigt fit to the observed data at \textit{T} = 2.0 K and show the
very good quality of the crystal with a full width half maximum
(FWHM) of 0.05$^o$. The decrease of the intensities as the
temperature is increased toward the bulk \textit{T}$_N$
demonstrates the magnetic character of this reflection, and is in
very good agreement with the N\'{e}el temperature obtained from
specific-heat measurements. The smooth decrease of the intensity
through \textit{T}$_N$ is a signature of a second-order-type
transition.

Figure~\ref{fig:Temp_Rh}b shows the energy dependence of the
$(\frac{1}{2},\frac{1}{2},9)$ magnetic reflection around the
Ce-L$_2$ absorption edge at \textit{T} = 2.0 K. A resonant
enhancement was observed at 1 eV above the absorption edges
revealing a result consistent with a dominant electric dipolar
character (\textit{E}1) of this transition (from 2\textit{p} to
5\textit{d} states). These results confirm the magnetic origin of
the $(\frac{1}{2},\frac{1}{2},l)$ reflections due to the existence
of an antiferromagnetic structure doubled in the basal plane.

To completely determine the magnetic structure of the system an
important aspect is the orientation of the magnetic ordered moment
with respect to the crystal lattice. While the magnetic wave
vector gives the relative orientation between the neighboring
spins, the direction of moment gives information about the
magnetic anisotropy (e.g. CEF effects) of the ordered spins. For
collinear magnetic structures and polarized incident photons, the
polarization dependence of the XRMS assumes a simple form for
dipolar resonances, and the intensities of magnetic Bragg peaks
are given by \cite{Hill_Acta1996}:

\begin{eqnarray} \label{eq:equation1}
I\propto\frac{1}{\mu^{*}sin(2\theta)}\left|\sum_{n}\textit{f}_{nE1}^{XRES}(\vec{k},\hat{\epsilon},\vec{k'},\hat{\epsilon'},\hat{z}_{n})e^{i\vec{Q}
\cdot \vec{R}_n}\right|^{2},
\end{eqnarray}

where $\mu^{*}$ is the absorption correction for asymmetric
reflections, 2$\theta$ is the scattering angle,
$\vec{Q}=\vec{k'}-\vec{k}$ is the wave-vector transfer, $\vec{k}$
and $\vec{k'}$ ($\hat{\epsilon}$ and $\hat{\epsilon'}$) are the
incident and scattered wave (polarization) vectors, respectively.
$\vec{R}_{n}$ is the position of the \textit{n}th resonant atom in
the lattice, e$^{i\vec{Q} \cdot \vec{R}_n}$ is the magnetic
structure factor, and finally $\hat{z}_{n}$ is the unit vector of
the moment direction of the Ce$^{3+}$ ions . The resonant
scattering amplitude contains both dipole (\textit{E}1) and
quadrupole (\textit{E}2) contributions. For the determination of
the magnetic structure we have used the second term of the
electric dipole transition (\textit{E}1) form factor
($\textit{f}_{nE1}^{XRES}$) which can be writen as:

\begin{eqnarray} \label{eq:equation2}
\textit{f}_{nE1}^{XRES} \propto \left[ \begin{array}{cc}
0 & \hat{k}\cdot\hat{z}_{n} \\
-\hat{k'}\cdot\hat{z}_{n} & (\hat{k}\times\hat{k'})\cdot\hat{z}_{n}\\
\end{array} \right]
\end{eqnarray}

\begin{eqnarray}\label{eq:equation3}
\propto \left[ \begin{array}{cc}
0 & z_1 cos \theta + z_3 sin \theta\\
- z_1 cos \theta + z_3 sin \theta & -z_2 sin (2\theta)\\
\end{array} \right],
\end{eqnarray}

where $\theta$ is the Bragg angle, $z_1$, $z_2$ and $z_3$ are the
components of the magnetic moment at the \textit{n}th site,
according to the commonly used geometry convention of Ref.
\onlinecite{Blume_PRB1988}. In this convention each term of the
matrix (Eq.~\ref{eq:equation2}) represents one polarization
channel: $\sigma$, $\pi$, $\sigma$' and $\pi$' describe the
incident (non-primed) and scattered (primed) photon polarizations.
To describe the orientation of the magnetic moment, $\eta$ is the
angle of the moment with relation of the \textit{ab}-plane and
$\psi$ the angle in the \textit{ab}-plane.

The magnetic structure of Cd-doped Ce$_2$RhIn$_{7.79}$Cd$_{0.21}$
was resolved  by comparing the intensities of the
$(\frac{1}{2},\frac{1}{2},\textit{l})$ (with \textit{l} = 4, 5, 6,
7, 8, 9 and 10) magnetic reflections with those expected using the
model given in Eqs. ~\ref{eq:equation1}
-~\ref{eq:equation3}\cite{Serrano_PRB74_2006,Hill_Acta1996,Cris_PRB2007,Granado_PRB69_2004,Mannix_EPJB2001,Cris_PhysB2009Ho}.
For this determination we used the data obtained for
($\sigma$-$\pi$') polarization channel measured with $\sigma$
incident photons in the energy of the Ce-L$_2$ edge and \textit{T}
= 2.0 K. For this channel the component of the matrix used was the
term -${\hat{k}'\cdot\hat{z}_{n}}$ of the Eq.~\ref{eq:equation2}.
Since the tetragonal structure of Cd-doped Ce$_2$RhIn$_8$ contains
two magnetic Ce$^{3+}$ ions per unit cell in the
\textit{c}-direction, this implies that two possibilities of an
AFM coupling can take place along the \textit{c}-axis: 1) sequence
+ + or 2) sequence + -, where the symbols + and - represent the
relative orientation of the magnetic moment of one Ce$^{3+}$ ion
with respect to the other. We checked the two sequences in our
model and the sequence + - gives the better results, in agreement
with the result obtained for pure Ce$_2$RhIn$_8$
\cite{Bao_PRB2001}.

\begin{figure}
\begin{center}
\vspace{-0.10cm}
\includegraphics[width=1.3\columnwidth,keepaspectratio]{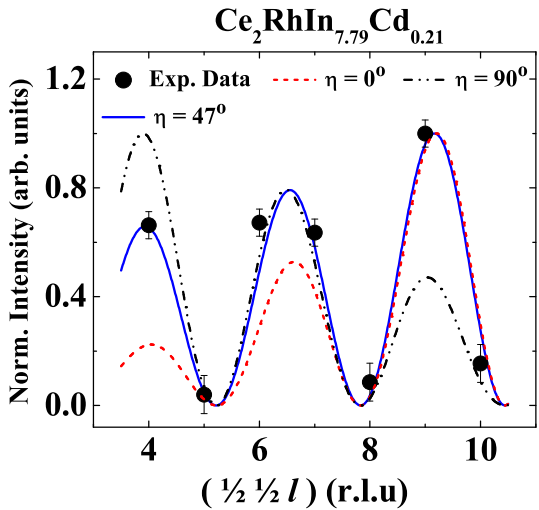}
\vspace{-2.0cm}
\end{center}
\caption{The \textit{l} dependence (in reciprocal lattice units)
of the normalized intensities of the magnetic peaks
$(\frac{1}{2},\frac{1}{2},\textit{l})$ measured at the energy of
the Ce-L$_2$ edge at \textit{T} = 2.0 K in the $\sigma$-$\pi$'
polarization channel. The solid line represents the best fit using
the model discussed in Eqs.~\ref{eq:equation1}-~\ref{eq:equation3}
for $\eta$ = 47$^o$ and $\psi$ = 45$^o$. The other lines represent
the model calculated for $\psi$ = 45$^o$ and $\eta$ = 0$^o$
(dashed line) and $\eta$ = 90$^o$ (dashed-dot line)
}\label{fig:Moment_Rh}
\end{figure}

The experimental intensities of the magnetic peaks
$(\frac{1}{2},\frac{1}{2},l)$ for Ce$_2$RhIn$_{7.79}$Cd$_{0.21}$
are compared to the model discussed above and displayed in
Figure~\ref{fig:Moment_Rh}. The best fit (solid line in
Fig.~\ref{fig:Moment_Rh}) using the model was obtained for $\eta$
= 47$^o$ $\pm$ 3$^o$ and $\psi$ = 45$^o$ $\pm$ 10$^o$.

The results in Fig.~\ref{fig:Moment_Rh} completely determine the
magnetic structure of the Ce$_2$RhIn$_{7.79}$Cd$_{0.21}$ sample.
The magnetic moment direction shows a slight evolution toward the
basal plane when compared with the pure Ce$_2$RhIn$_8$ where the
moment direction at 52$^o$ to the basal plane. Those results
indicate that Cd-doping also subtly changes the single ion
anisotropy of the CEF GS which mainly determines the direction of
the ordered moment. Analyzing these results in a more general way
one can compare our results with those found for the In-based
R$_{m}$M$_{n}$In$_{3m+2n}$ (M = Co, Rh or Ir; M = 1, 2; R =
rare-earth) series. In a recent theoretical work using a mean
field model including an isotropic first-neighbors (RKKY)
interaction and the tetragonal CEF\cite{Pagliuso_JAP2006}, it was
demonstrated that, for realistic values of the interactions and
CEF parameters, it is possible to determine qualitatively the
direction of the ordered moments and the behavior of the ordering
temperature for these series. Following the trend obtained for the
In-based R$_{m}$M$_{n}$In$_{3m+2n}$ when the N\'{e}el temperature
is higher the magnetic moment direction tends towards the
\textit{c}-axis, considering dominant CEF effects. Therefore,
although the Cd-doping induces some changes in the CEF parameters,
our results show that the direction of the ordered moment tend to
rotate to the \textit{ab}-plane even with an increase of the
\textit{T}$_N$ for a Cd-doped sample, strongly suggesting that the
dominant effect to raise \textit{T}$_N$ in these systems is the
electronic tuning induced by the Cd-doping. The indication that
the tetragonal CEF is changing as a function of Cd-doping in these
series may explain the behavior of \textit{T}$_{MAX}$ in
electrical resistivity measurements. If this subtle change in the
CEF parameters leads to an increase of the energy level of the
first excited doublet it may cause an increase in
\textit{T}$_{MAX}$. However, further experiments to obtain direct
evidence of this change of CEF scheme as a function of Cd-doping
is necessary to confirm this claim.

The NMS experiments were performed to study both Cd-doped
Ce$_2$RhIn$_8$ and Ce$_2$IrIn$_8$ for \textit{x} = 0.21 of Cd
concentration. The Rh member was studied by NMS to determine the
effective magnetic moment of the Ce$^{3+}$ ions in low-\textit{T}.
The experiments were performed  with an incident energy of 35 meV
with no absorption corrections based on the fact that at 35 meV
the neutron penetration length is about 2.0 mm, which is larger
than the thickness of the sample. Indeed we measured rocking
curves for different domains and no significant changes on the
intensities were noticed. Magnetic peaks of the
Ce$_2$(Rh,Ir)In$_{7.79}$Cd$_{0.21}$ were observed at temperatures
below $T_N$ = 4.8 K and 3.8 K, respectively for Rh and Ir samples,
at reciprocal lattice points forbidden for nuclear scattering and
consistent with an antiferromagnetic structure with propagation
vector $(\frac{1}{2},\frac{1}{2},0)$. These results show that the
Cd-doped Ce$_2$IrIn$_8$ sample presents the same propagation
vector found for pure and Cd-doped Ce$_2$RhIn$_8$ samples.

\begin{figure}
\begin{center}
\vspace{-0.10cm}
\includegraphics[width=1.3\columnwidth,keepaspectratio]{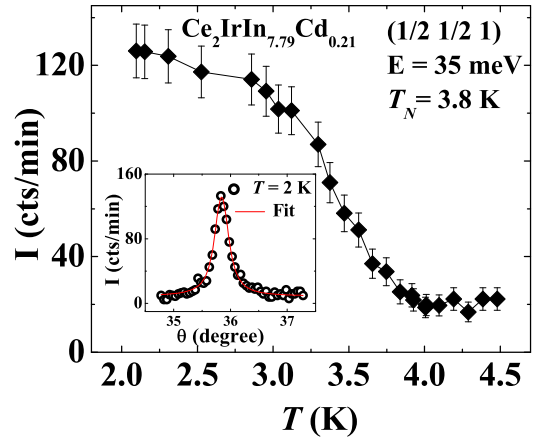}
\vspace{-2.0cm}
\end{center}
\caption{Temperature dependence of the integrated intensities of
the $(\frac{1}{2},\frac{1}{2},1)$ magnetic reflection measured
during heating the sample in the temperature range between
\textit{T} = 2.0 K and \textit{T} = 4.5 K for
Ce$_2$IrIn$_{7.79}$Cd$_{0.21}$ for 35 meV. The inset shows the
$\theta$-scan of the same reflection for \textit{T} = 2.0 K where
the continuous line is a Voigt fit to the observed
data.}\label{fig:Temp_Ir}
\end{figure}

Figure~\ref{fig:Temp_Ir}a displays the temperature dependence of
the integrated intensity, of the $(\frac{1}{2},\frac{1}{2},1)$
magnetic reflection at an incident energy of 35 meV and measured
between \textit{T} = 2.0 K and \textit{T} = 4.5 K while warming
the sample. The inset shows an experimental $\theta$-scan for
\textit{T} = 2.0 K. The continuous line is a Voigt fit to the
observed data at \textit{T} = 2.0 K and shows a FWHM of 0.2$^o$, a
value larger than the one obtained for X-ray diffraction because
of the larger divergence of the neutron beam compared with X-rays.
The decrease of the intensity as the temperature is increased
toward the bulk \textit{T}$_N$ indicates the magnetic character of
this reflection, and are in very good agreement with the N\'{e}el
temperature obtained from specific-heat measurements. The smooth
decrease of the intensity is a signature of a second-order-type
transition, found also for Ce$_2$IrIn$_{7.79}$Cd$_{0.21}$ sample.

To completly determine the magnetic structure of
Ce$_2$IrIn$_{7.79}$Cd$_{0.21}$ it is necessary to determine the
magnetic moment orientation of the Ce$^{3+}$ ion in relation to
the crystallographic axis. Integrated intensities of magnetic
Bragg peaks were determined using a Voigt fit in the
$\theta$-scans. These magnetic peaks were first normalized using
the structural Bragg peaks (00\textit{l}) for \textit{l} = 1, 2,
3, 4, and 7; (11\textit{l}) for \textit{l} =1, 2, 3, 4, 5, and 6
and (22\textit{l}) for \textit{l} = 0, 1, 2, 3, and 6. In barn
units the magnetic cross section for a collinear magnetic
structure using unpolarized neutrons is given by
\cite{Squires,Shirane,Bao_PRB2001}:

\begin{eqnarray} \label{eq:equation4}
\sigma(Q)=\left(\frac{\gamma
r_{0}}{2}\right)^{2}\left|M\right|^{2}{\displaystyle
\left|f(Q)\right|^{2}
\sum_{\mu,\nu}(\delta_{\mu,\nu}-\hat{Q}_{\mu}\hat{Q}_{\nu})}{\displaystyle
F_{\mu}^{*}(Q)F}_{\nu}(Q)
\end{eqnarray}

where ($\gamma$\textit{r}$_0$/2)$^2$ = 0.07265 b/$\mu_{B}^{2}$, M
is the effective magnetic moment of the Ce$^{3+}$ ion,
\textit{f}(Q) is the Ce$^{3+}$ magnetic form factor \cite{Blume},
and F$_\mu$(Q) is the $\mu^th$ cartesian component of magnetic
structure factor per Ce$_2$IrIn$_8$. The calculations were made
considering the average of the possible domains and the result is
given by \cite{Bao_PRB2001}:

\begin{eqnarray} \label{eq:equation5}
\sigma(Q)=\left(\frac{\gamma
r_{0}}{2}\right)^{2}\left|M\right|^{2}{\displaystyle
\left|f(Q)\right|^{2}\left\langle
1-(\hat{Q}\cdot\hat{z_n})^{2}\right\rangle
}\left|F_{M}(Q)\right|^{2}
\end{eqnarray}

where F$_M$(Q) is the magnetic form factor calculated for the two
Ce$^{3+}$ ions of the unitary cell along the \textit{c}-axis,
$\hat{z_n}$ is the unit vector of the magnetic moment, and the
average, $\left\langle 1-(\hat{Q}\cdot\hat{z_n})^{2}\right\rangle$
is over the magnetic domains \cite{Bao_PRB2001}.

The NMS technique does not allow the moment direction
determination in the \textit{ab}-plane due to the square symmetry
\cite{Squires,Shirane}, only the moment direction relative to the
\textit{c}-axis is allowed. Considering the magnetic moment with
an arbitrary orientation in relation to the \textit{c}-axis, there
are in general 16 magnetic domains with tetragonal symmetry, for
an equal population of domains, the average term of the
Eq.~\ref{eq:equation5} could be written as\cite{Bao_PRB2001}:

\begin{eqnarray} \label{eq:equation6}
\left\langle 1-(\hat{Q}\cdot\hat{z_n})^{2}\right\rangle
=1-\frac{\cos^{2}\alpha\,\cos^{2}\eta+2\,\sin^{2}\alpha\,\sin^{2}\eta}{2}
\end{eqnarray}
here $\alpha$ is the angle of $\vec{Q}$ in relation to the basal plane and $\eta$
is the angle between the moment direction and the basal plane.

\begin{figure}
\begin{center}
\vspace{-0.10cm}
\includegraphics[width=1.3\columnwidth,keepaspectratio]{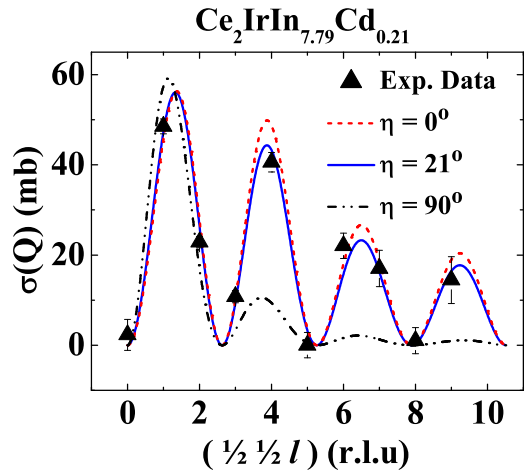}
\vspace{-2.0cm}
\end{center}
\caption{The \textit{l} dependence (in reciprocal lattice units)
of the $\sigma$(Q) of the magnetic peaks
$(\frac{1}{2},\frac{1}{2},\textit{l})$ measured at the energy of
35 meV at \textit{T} = 2 K. The solid line represents the best fit
using the model discussed in Eqs.~\ref{eq:equation4}
-~\ref{eq:equation6} for $\eta$ = 21$^o$ and M = 0.22 $\mu_{B}$.
The other lines represent the model calculated for $\eta$ = 0$^o$
(dashed line) and $\eta$ = 90$^o$ (dashed-dot
line).}\label{fig:Moment_Ir}
\end{figure}

Figure~\ref{fig:Moment_Ir} shows the \textit{l} dependence of the
experimental $(\frac{1}{2},\frac{1}{2},\textit{l})$ magnetic
intensities compared with the magnetic cross section $\sigma$(Q)
in mbarn units calculated using the model discussed in
Eqs.~\ref{eq:equation4} -~\ref{eq:equation6}. The solid line in
Fig.~\ref{fig:Moment_Ir} displays the best fit obtained for $\eta$
= 21$^o$ $\pm$ 5$^o$ and the ordered magnetic moment at 2 K was
determined as M = 0.4(5) $\mu_{B}$ per Ce. The model calculated
for $\eta$ = 0$^o$ is represented by the dashed line while the
dashed-dot line shows the model to $\eta$ = 90$^o$.

The results in Fig.~\ref{fig:Moment_Ir} completely determine the
magnetic structure of Ce$_2$IrIn$_{7.79}$Cd$_{0.21}$ sample. The
magnetic moment direction of the doped compound shows a large
evolution toward to the basal plane when compared with the
Ce$_2$RhIn$_{7.79}$Cd$_{0.21}$ where the moment direction lies at
47$^o$ of the basal plane.

\begin{table}[htb!]
\caption{Magnetic Bragg intensities, $\sigma_{obs}$, (mbarn)
observed at 1.6 K with E = 35 meV. The theoretical intensities,
$\sigma_{calc}$ (mbarn), were calculated using
Eqs.~\ref{eq:equation4} -~\ref{eq:equation6} with $\eta$ = 47$^o$
and M = 0.9(1) $\mu_B$ per Ce.} \label{T1}\centering
\begin{tabular}{ccc}
\hline \hline
$\vec{Q}$ & $\sigma_{exp}$ & $\sigma_{calc}$ \\
 \hline
$(\frac{1}{2},\frac{1}{2},0)$ & 0.1(2) & 0 \\
  $(\frac{1}{2},\frac{1}{2},1)$ & 138(2) & 134 \\
  $(\frac{1}{2},\frac{1}{2},2)$ & 61(1) & 56 \\
  $(\frac{1}{2},\frac{1}{2},3)$ & 21(1) & 19 \\
  $(\frac{1}{2},\frac{1}{2},4)$ & 66(1) & 69 \\
  $(\frac{1}{2},\frac{1}{2},5)$ & 3(2) & 4 \\
  $(\frac{1}{2},\frac{1}{2},6)$ & 39(1) & 30 \\
  $(\frac{1}{2},\frac{1}{2},7)$ & 38(1) & 28 \\
\hline \hline
\end{tabular}
\end{table}

NMS experiments were performed also in
Ce$_2$RhIn$_{7.79}$Cd$_{0.21}$ to complement the study made with
XRMS and to determine the effective magnetic moment M of this
compound in low-\textit{T}. Table I shows the observed magnetic
integrated intensities of the
$(\frac{1}{2},\frac{1}{2},\textit{l})$ (l = 0, 1, 2, 3, 4, 5, 6
and 7) magnetic reflections normalized by the nuclear peaks
(00\textit{l}) for \textit{l} = 1, 2, 3, 4, and 7; (11\textit{l})
for \textit{l} =1, 2, 3, 4, 5, and 6 and (22\textit{l}) for
\textit{l} = 0, 1, 2, 3, and 6; the theoretical cross section was
calculated using Eqs.~\ref{eq:equation4} -~\ref{eq:equation6} and
the best fit the data is $\eta$ = 47(5)$^o$ and an ordered
magnetic moment M = 0.9(2) $\mu_B$. The result of the magnetic
moment direction found for Ce$_2$RhIn$_{7.79}$Cd$_{0.21}$ sample
is in very good agreement with the result found by means of the
XRMS, showing the consistency of the two experimental methods. The
novelty here is the ordered magnetic moment for Ce$^{3+}$ ion in
low-T equal to 0.9(2) $\mu_B$ that is a value larger than the
value found for pure Ce$_2$RhIn$_8$, also consistently with the
fact that Cd doping is favoring the AFM order in this system.

\section{CONCLUSIONS}

The low temperature physical properties of Cd-doped Ce$_2$MIn$_8$
(M = Rh, Ir) single crystalline samples have been investigated.
The results of the susceptibility, heat-capacity and resistivity
as a function of temperature revealed an enhancement of the
N\'{e}el temperature from 2.8 K for the pure compound to 4.8 K for
the Cd-doped sample for M = Rh. For M = Ir an ordered AFM state at
low Cd concentration and a subsequent evolution of the N\'{e}el
temperature up to 3.8 K is found. This confirms the trend observed
for the 115 compounds that the Cd-doping favors AFM ordering in
these series and suggests that Cd is acting as an electronic
tuning agent changing the local density of states that acts
fundamentally on the GS behavior. However there exist experimental
evidences that Cd doping is affecting the tetragonal CEF
parameters in these compounds. This is reflected in the tendency
of the ordered moment to rotate to the $ab$-plane in Cd-doped
samples. An evolution of the CEF scheme may also help to
understand the behavior of \textit{T}$_{MAX}$ that increases as a
function of Cd concentration. The increase of this parameter is
not expected in a scenario where Cd is  just acting as an
electronic tuning agent. Further experiments like inelastic
neutron scattering should be performed aiming to clarify the
function of the Cd doping in the microscopic behavior on these
systems. The XRMS and NMS experiments showed that Cd-doped
Ce$_2$(Rh,Ir)In$_8$ has a commensurate magnetic order with
propagation vector $\vec{\varepsilon}$ =
$(\frac{1}{2},\frac{1}{2},0)$ identical to the pure Ce$_2$RhIn$_8$
compound, indicating that Cd-doping does not alter the sign of the
relative magnetic interaction between the neighboring Ce spins,
although it increases the magnitude of the average magnetic
interaction between them as it can be seen by the increase of the
ordered magnetic moment in the Cd-doped compounds.

\begin{acknowledgments}
This work was supported by FAPESP (SP-Brazil), CNPq (Brazil) and
CAPES (Brazil). The staff at the ID-20 beam line and BT-9
instrument are gratefully acknowledged for providing an
outstanding scientific environment during these experiments.
\end{acknowledgments}

\bibliography{basename of .bib file}

\begin{thebibliography}{99}


\bibitem{Monthoux}See for instance: P. Monthoux and G. G. Lonzarich, Phys. Rev. B, 59, 14598
(1999).
\bibitem{Tuson} T. Park, F. Ronning, H. Q. Yuan, M. B. Salamon, R. Movshovich, J. L. Sarrao and J. D. Thompson, Nature \textbf{440}, 65 (2006).

\bibitem{Hegger_PRL2000} H. Hegger, C. Petrovic, E. G. Moshopoulou, M. F. Hundley, J. L. Sarrao,
Z. Fisk, and J. D. Thompson, Phys. Rev. Lett \textbf{84}, 4986
(2000).
\bibitem{Petrovic_JPCM2001} C. Petrovic, P. G. Pagliuso, M. F. Hundley, R. Movshovich,
J. L. Sarrao, J. D. Thompson, Z. Fisk, and P. Monthoux, J. Phys.:
Condens. Matter, \textbf{13}, L337 (2001).


\bibitem{PG_PhysB_320_2002} P. G. Pagliuso, N. J Curro, N. O. Moreno, M.  F. Hundley, J. D. Thompson,
J. L. Sarrao and Z. Fisk, Physica B \textbf{320}, 370-375 (2002).
\bibitem{PG_PRB2002} P. G. Pagliuso, N. O. Moreno, N. J. Curro, J. D. Thompson, M. F. Hundley,
J. L. Sarrao, and Z. Fisk, Phys. Rev. B \textbf{66}, 054433 (2002).
\bibitem{Pagliuso_PRB2001} P. G. Pagliuso, C. Petrovic, R. Movshovich, D. Hall, M. F. Hundley,
J. L. Sarrao, J. D. Thompson and Z. Fisk, Phys. Rev. B, 64
100503(R)(2001).
\bibitem{Zheeng} Guo-qing Zheng \textit{et al.}, Phys. Rev. B \textbf{70}, 014511 (2004).
\bibitem{Tuson2} T. Park, V. A. Sidorov, F. Ronning, J.-X. Zhu, Y. Tokiwa, H. Lee, E. D. Bauer, R. Movshovich, J. L. Sarrao and J. D. Thompson, Nature \textbf{456}, 366 (2008).
\bibitem{Nicklas2} M. Nicklas, V. A. Sidorov, H. A. Borges, P. G. Pagliuso, C. Petrovic, Z. Fisk, J. L. Sarrao and J. D. Thompson, Phys. Rev. B \textbf{67}, 020506(R) (2003).
\bibitem{PG_PhysB_312_2002} P.G. Pagliuso, R. Movshovich, A. D. Bianchi, M. Nicklas, N. O. Moreno, J. D. Thompson, M.F. Hundley, , J.L. Sarrao and Z. Fisk,
Physica B \textbf{312-313}, 129-131 (2002).

\bibitem{Ferreira_PRL2008} L. Mendon\c{c}a-Ferreira, T. Park, V. Sidorov, M. Nicklas, E. M. Bittar,
R. Lora-Serrano, E. N. Hering, S. M. Ramos, M. B. Fontes, E. Baggio-Saitovich, Hanoh Lee, J. L. Sarrao, J. D. Thompson and P. G.
Pagliuso, Phys. Rev. Lett. \textbf{101}, 017005 (2008).

\bibitem{ericPRB} E. D. Bauer \textit{et al.}, Phys. Rev. B \textbf{73}, 245109 (2006).



\bibitem{Ramos_PhysB2005}S. M. Ramos, M. B. Fontes, A. D. Alvarenga, E. Baggio-Saitovitch, P. G. Pagliuso, E. D. Bauer, J. D. Thompson, J. L. Sarrao, M. A.
Continentino, Physica B \textbf{359-361}, 398-400 (2005).
\bibitem{Daniel} M. Daniel, E. D. Bauer, S. -W, Han, C. H. Booth, A. L. Cornelius, P. G. Pagliuso, J. L. Sarrao, Phys. Rev. Lett. {\bf 95}, 016406 (2005).
\bibitem{Pham_PRL2006} L. D. Pham, Tuson Park, S. Maquilon, J. D. Thompson and Z. Fisk, Phys. Rev. Lett. \textbf{97}, 056404 (2006).
\bibitem{Nicklas} M. Nicklas, O. Stockert, T. Park, K. Habicht, K. Kiefer,
L.D. Pham, J.D. Thompson, Z. Fisk, F. Steglich, Phys. Rev. B
\textbf{70}, 014511 (2004).
\bibitem{Urbano_PRL2007} R. R. Urbano, B.-L. Young, N. J. Curro, J. D. Thompson, L. D. Pham and Z.
Fisk, Phys. Rev. Lett., \textbf{99}, 146402 (2007)

\bibitem{pagliuso3} P. G. Pagliuso \textit{et al.}, Phys. Rev. B \textbf{66}, 054433 (2002).


\bibitem{Hering_PhysB2008} E. N. Heringa, H. A. Borges, S. M. Ramos, M. B. Fontes, E. Baggio-Saitovich,
E. M. Bittar, L. Mendon\c{c}a Ferreira, R. Lora-Serrano, C. Adriano, P.G. Pagliuso, J.L. Sarrao, J.D. Thompson, Physica B,
\textbf{403}, 780-782 (2008).
\bibitem{Moreno1} N. O. Moreno, M. F. Hundley, P. G. Pagliuso, R. Movshovich, M. Nicklas, J. D. Thompson,
J. L. Sarrao and Z. Fisk, Physica B \textbf{312-313}, 274 (2002).
\bibitem{Ce2PdIn8} D. Kaczorowski, A. P. Pikul, D. Gnida and V. H. Tran, Phys. Rev. Lett. \textbf{103}, 027003 (2009).

\bibitem{Bao_PRB2001}  Wei Bao, P.G. Pagliuso, J.L. Sarrao, J. D. Thompson, Z. Fisk, and J. W. Lynn, Phys. Rev. B \textbf{64}, 020401(R) (2001).
\bibitem{morris} G. D. Morris, R. H. Heffner, N. O. Moreno, P. G. Pagliuso, J. L. Sarrao, S. R. Dunsiger, G. J. Nieuwenhuys, D. E. MacLaughlin and O. O. Bernal, Phys. Rev. B \textbf{69}, 214415 (2004).

\bibitem{PG_PRB63_2001}  P. G. Pagliuso, J. D. Thompson, M. F. Hundley, J. L. Sarrao and Z. Fisk, Phys. Rev. B \textbf{63}, 054426 (2001).
\bibitem{Fisk_Handbook1989} Z. Fisk and J. P. Remeika, \emph{Handbook on the Physics and Chemistry of Rare
  Earths} (Elsevier, North-Holland, 1989) vol. 12, p. 53.
\bibitem{PG_PRB62_2000} P. G. Pagliuso, J. D. Thompson, M. F. Hundley and J. L.
Sarrao, Phys. Rev. B \textbf{62}, 12266 (2000).
\bibitem{Pagliuso_JAP2006} P. G. Pagliuso, D. J. Garcia, E. Miranda, E. Granado, R. Lora Serrano, C. Giles, J. G. S. Duque, R. R. Urbano, C. Rettori, J. D. Thompson, M. F. Hundley and J. L. Sarrao, J. Appl. Phys. \textbf{99}, 08P703 (2006).
\bibitem{Serrano_PRB74_2006} R. Lora-Serrano, C. Giles, E. Granado, D. J. Garcia, E. Miranda, O. Agüero, L. Mendon\c{c}a-Ferreira, J. G. S. Duque, and P. G.
Pagliuso, Phys. Rev. B \textbf{74}, 214404 (2006).
\bibitem{Hieu_JMMM2007} N.V. Hieu, H. Shishido, H. Nakashima, K. Sugiyama, R. Settai, T. Takeuchi, T.D. Matsuda, Y. Haga, M. Hagiwara, K. Kindo, Y. Onuki, J.
Magn. Magn. Mater. \textbf{310}, 1721 (2007).
\bibitem{Hudis_JMMM2006} J. Hudis, R. Hu, C. L. Broholm, V. F. Mitrovi\'{c}, C.
Petrovic, J. Magn. Magn. Mater. \textbf{307}, 301 (2006).
\bibitem{Doniach} S. Doniach, in \textit{Valence Instabilities and Related Narrow
Band Phenomena}, edited by R. D. Parks (Plenum, New York, 1977),
p. 169.

\bibitem{Eva1} E. G. Moshopoulou, Z. Fisk, J. L. Sarrao and J. D.Thompson, J. Solid State Chem. \textbf{158}, 25 (2001).
\bibitem{Eva2} E. G. Moshopoulou, R. M. Ibberson, J. L. Sarrao, J. D. Thompson and Z. Fisk, Acta Crystallogr. B \textbf{62}, 173 (2006).

\bibitem{Cris_PhysB2009}C. Adriano, C. Giles, L. Mendon\c{c}a-Ferreira, F. de Bergevin, C.Mazzoli, L.Paolasini,
Z. Fisk, P.G.Pagliuso, Physica B \textbf{404}, 3014 (2009).
\bibitem{Hill_Acta1996} J. P. Hill and D. F. McMorrow, Acta Crystallogr., Sect. A: Found.
Crystallogr. \textbf{A52}, 236 (1996).
\bibitem{Blume_PRB1988} M. Blume and D. Gibbs, Phys. Rev. B \textbf{37}, 1779 (1988).
\bibitem{Cris_PRB2007} C. Adriano, R. Lora-Serrano, C. Giles, F. de Bergevin, J. C. Lang, G. Srajer, C. Mazzoli,
L. Paolasini, and P. G. Pagliuso, Phys. Rev. B \textbf{76}, 104515 (2007).
\bibitem{Granado_PRB69_2004} E. Granado, P. G. Pagliuso, C. Giles, R. Lora-Serrano, F. Yokaichiya and J. L. Sarrao, Phys. Rev. B \textbf{69}, 144411 (2004).
\bibitem{Mannix_EPJB2001} D. Mannix, P. C. de Camargo, C. Giles, A. J. A. de Oliveira,
F. Yokaichiya and C. Vettier, Eur. Phys. J. B \textbf{20}, 19-25 (2001).
\bibitem{Cris_PhysB2009Ho} C. Adriano, C. Giles, L. N. Coelho, G. A. Faria, P. G. Pagliuso, Physica B \textbf{404}, 3289 (2009).
\bibitem{Squires}G. L. Squires, \emph{Introduction to the Theory of Thermal Neutron
Scattering} (Dover Publications, INC. Mineola, New York, 1996).
\bibitem{Shirane} G. Shirane, S. M. Shapiro, and J. M. Tranquada, \emph{Neutron Scattering with a Triple-Axis Spectrometer,
Basic Techniques} (Cambrige University Press, Cambridge, 2002)
\bibitem{Blume} M. Blume, A. J. Freeman, and R. E. Watson, J. Chem. Phys. \textbf{37},
1245 (1962).

\end{thebibliography}

\end{document}